\documentclass[cits]{PoS}

\title{
\vspace*{-6.3cm}
\begin{flushright}
{\scriptsize KA-TP-15-2014; SFB/CPP-14-31}
\end{flushright}
\vspace*{6.8cm}
A theoretical review of triple Higgs coupling studies at the
  LHC in the Standard Model}
\ShortTitle{$HHH$ coupling a the LHC: a theory review}

\author{\speaker{Julien Baglio}\\
  Institut f\"{u}r Theoretische Physik, Karlsruhe Institute of
  Technology (KIT),\\
  Wolfgang-Gaede Strasse 1, Karlsruhe D-76131 (Germany)\\
  E-mail: \email{julien.baglio@kit.edu}}

\abstract{
  After the discovery of a Higgs boson at the LHC, the next important
  step is to measure its couplings to fermions and bosons to unravel
  its true nature. In order to ultimately test the shape of the scalar
  potential that triggers the electroweak symmetry breaking, it is
  crucial to measure the triple Higgs coupling at the LHC. We then
  review the theoretical predictions of the main Standard Model Higgs
  pair production mechanisms that are needed for such a measurement
  and present the latest developments in the phenomenological analyses
  in view of a high luminosity LHC.}
      
\FullConference{XXII. International Workshop on Deep-Inelastic
  Scattering and Related Subjects,\\
  28 April - 2 May 2014\\
  Warsaw, Poland}

\begin{document}

\section{Introduction}

After the discovery of a bosonic particle in 2012 at CERN
~\cite{Aad:2012tfa} which has the properties of a Higgs
boson~\cite{Englert:1964et}, the next 
important task is to perform a detailed study of this boson to pin
down its exact nature. The LHC data in 2013 seem to favor a Standard
Model (SM) Higgs boson so far~\cite{ATLAS:conf}, but there is still
room left for a beyond-the-SM (BSM) interpretation. A measurement of
the Higgs boson self-couplings would allow for the reconstruction of
the scalar potential triggering the electroweak symmetry breaking
(EWSB) mechanism and is the ultimate test of the SM.

After EWSB, the scalar potential contains triple and quartic Higgs
couplings. It has been shown in the last decade that the quartic Higgs
coupling is not accessible at current of foreseen collider energies
of order 100 TeV~\cite{Plehn:2005nk}, leading to the focus on the
triple Higgs coupling which can be probed through the production of a
Higgs boson pair.
The early studies focused first on leptonic
colliders~~\cite{Boudjema:1995cb,Djouadi:1999gv} before the first
study at the LHC which gave the theoretical predictions for the main
production mechanisms~\cite{Djouadi:1999rca}. A comprehensive
analysis of the $b\bar{b}\gamma\gamma$ search channel, including a fit
to the $m_{HH}$ distributions, stated later on that excluding a
vanishing triple Higgs coupling would be possible at the LHC with a
very high luminosity of 6 ab$^{-1}$~\cite{Baur:2002rb}.

This review will present the recent improvements in the theoretical
predictions of the main production mechanisms and in the
phenomenological analyses compared to these early studies. Only the SM
case will be presented but numerous BSM studies have also been
performed.

\section{SM Higgs boson pair production at the LHC}

\begin{figure}
\begin{center}
\includegraphics[scale=0.5]{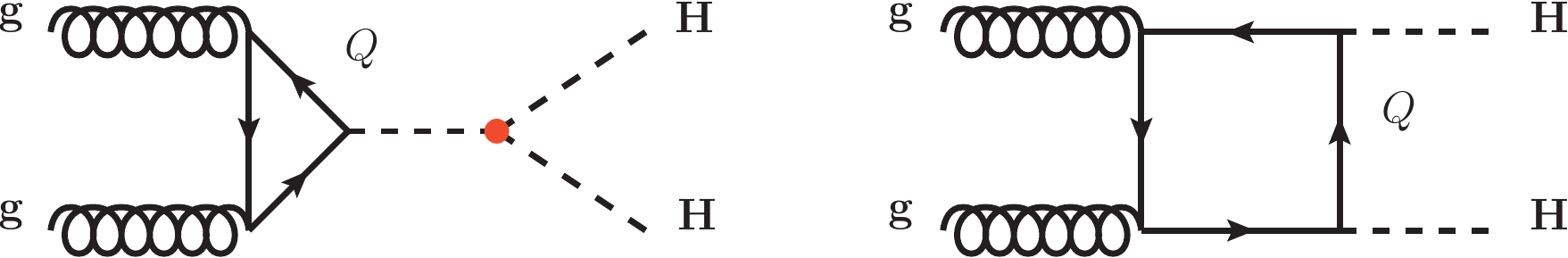}

\vspace{4mm}\includegraphics[scale=0.5]{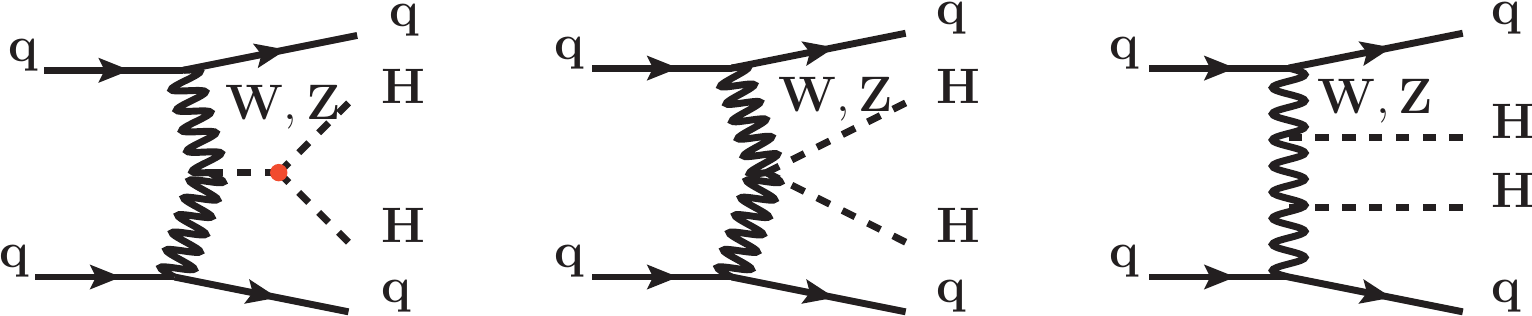}
\end{center}
\vspace{-5mm}\caption{Generic Feynman diagrams contribution to gluon
  fusion Higgs pair production (up) and VBF production (down). The
  triple Higgs coupling is highlighted in red.}
\label{fig:diagram}
\end{figure}

The main production channels for a Higgs boson pair follow the same
pattern as for single Higgs production. The main channel is the gluon fusion
production which is known up to next-to-next-to-leading order (NNLO) in
QCD in an effective field theory (EFT) approach~\cite{deFlorian:2013uza},
then followed by the vector boson fusion (VBF) channel known exactly
up to next-to-leading order (NLO) in
QCD~\cite{Baglio:2012np,Frederix:2014hta} and even up to NNLO in a
structure function approach~\cite{Liu-Sheng:2014gxa}, see
Fig.~\ref{fig:diagram} for 
generic Feynman diagrams. The two other channels are of less
importance, the double Higgs-strahlung known up to NNLO in
QCD~\cite{Baglio:2012np} and the associated production with a
top-antitop pair known up to NLO in QCD~\cite{Frederix:2014hta}.

\begin{figure}
\begin{center}
\begin{minipage}[c]{7cm}
\hspace{6mm}\includegraphics[scale=0.55]{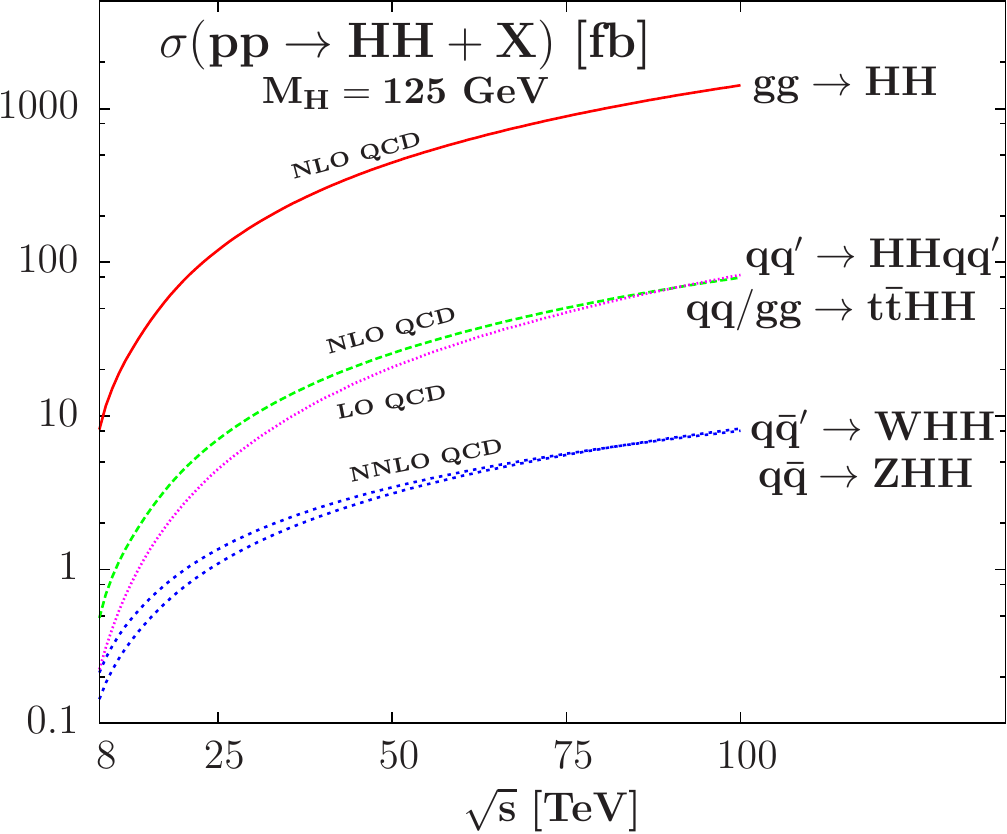}
\end{minipage}
\begin{minipage}[c]{7cm}
\vspace{-3mm}\includegraphics[scale=0.38]{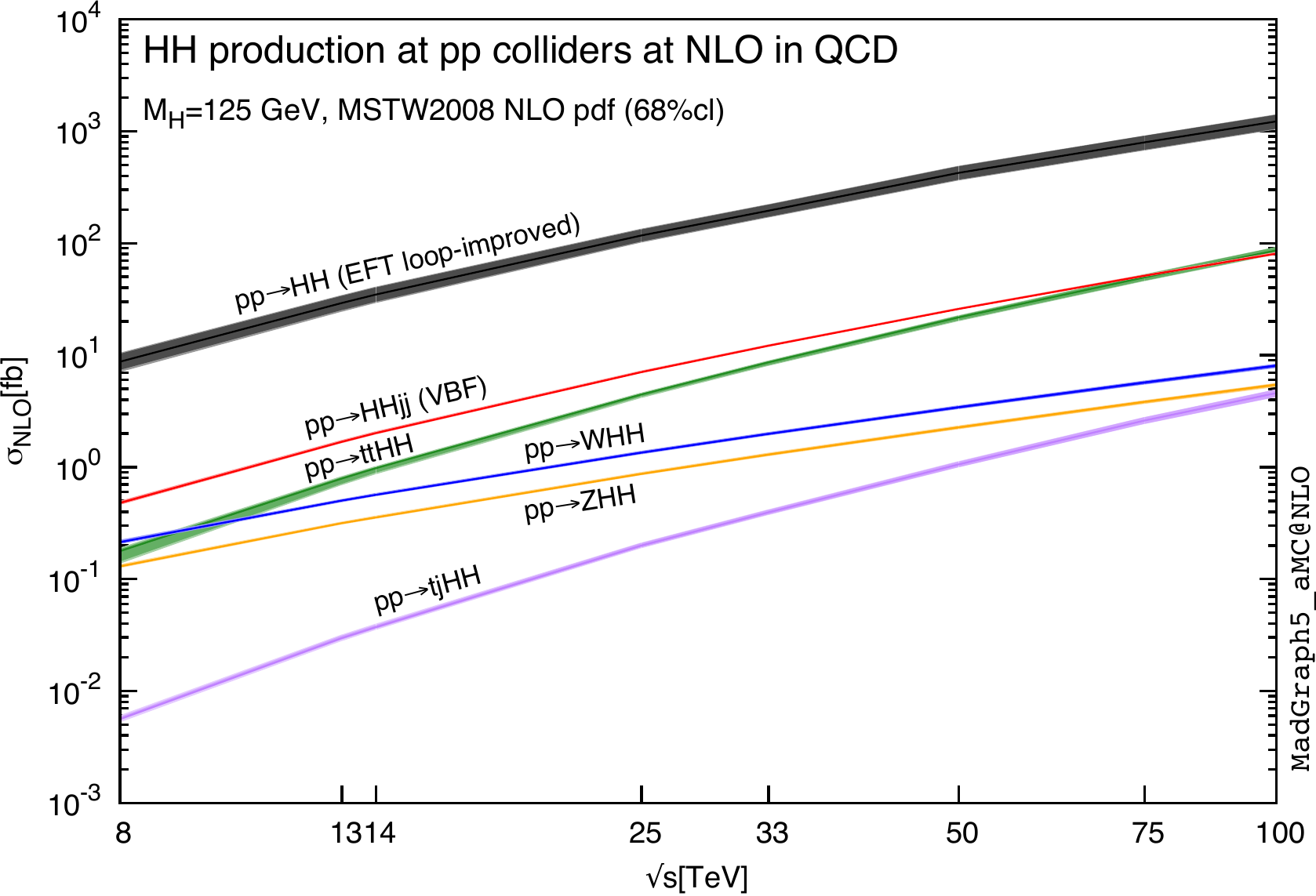}
\end{minipage}
\end{center}
\vspace{-5mm}\caption{The total hadronic cross section of the main
  production channels of a Higgs boson pair $HH$ (in fb) as a function
  of the center-of-mass-energy $\sqrt{s}$ (in TeV). Taken from
  Ref.~\cite{Baglio:2012np} (left) and Ref.~\cite{Frederix:2014hta}
  (right).}
\label{fig:main}
\end{figure}

The pattern of these four channels is depicted in Fig.~\ref{fig:main}
where the total cross section is presented as a function of the
center-of-mass energy. The important common feature
of all these channels is the smallness of the cross sections: compared
to single Higgs boson production they are three orders of magnitude
smaller and that explains how challenging the measurement of Higgs
boson pair production at the LHC is. A high luminosity will be
required to perform such a measurement.

\subsection{The gluon fusion channel}

\begin{figure}
\begin{minipage}[c]{5cm}
\includegraphics[width=4.7655cm,height=3.8115cm]{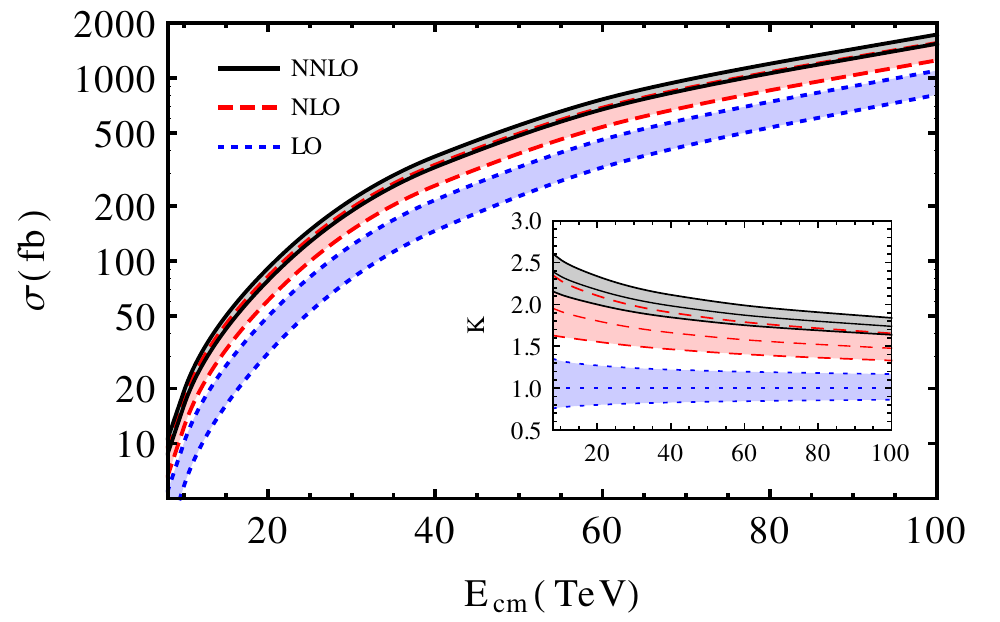}
\end{minipage}
\begin{minipage}[c]{5cm}
\includegraphics[scale=0.45]{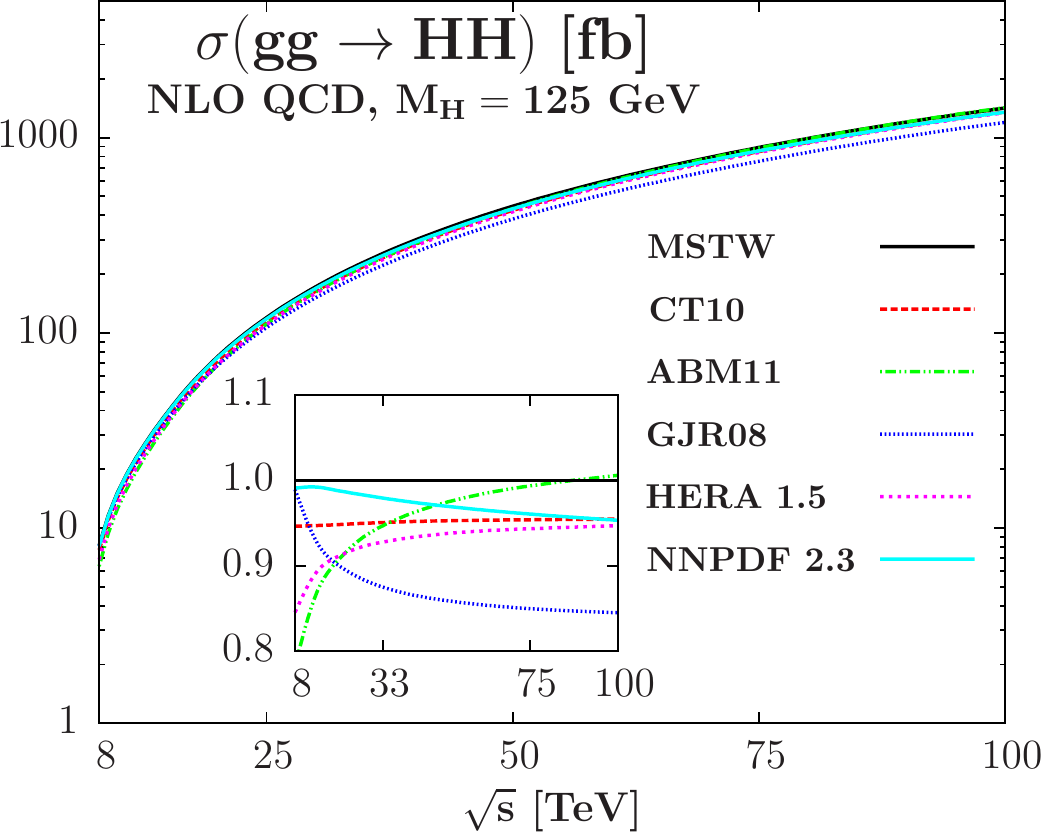}
\end{minipage}
\begin{minipage}[c]{4.5cm}
\includegraphics[scale=0.45]{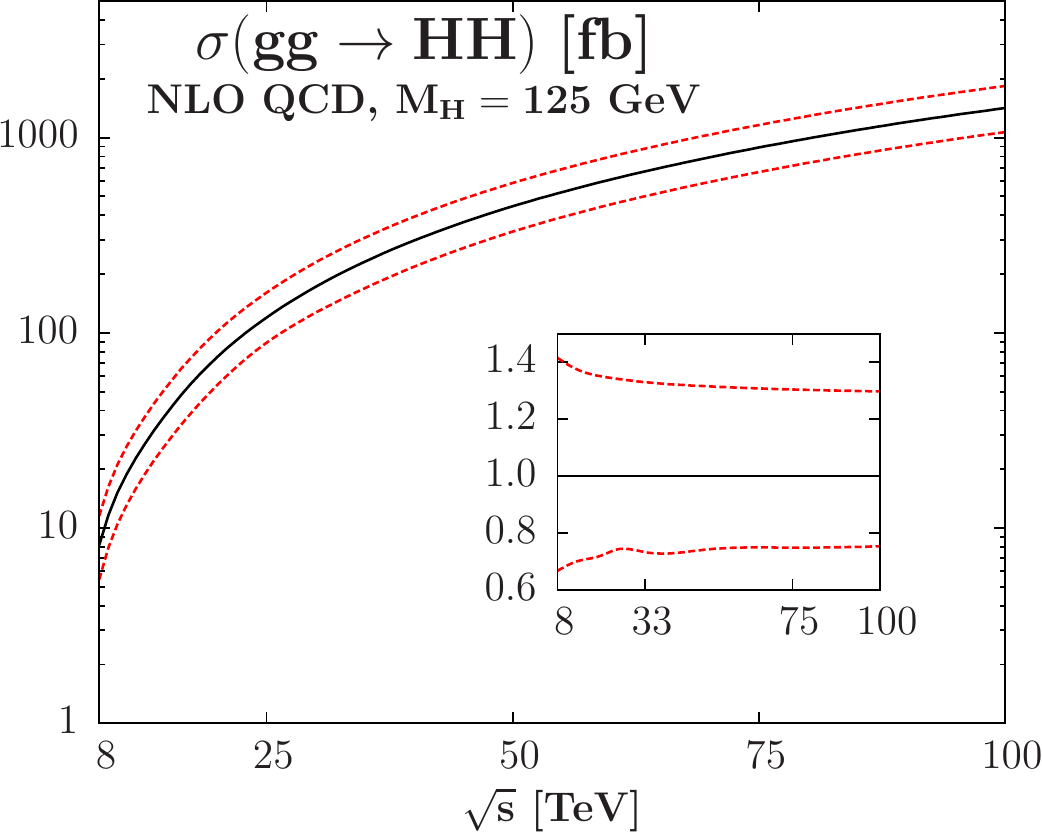}
\end{minipage}
\caption{Left: the total hadronic cross section $\sigma(gg\to HH)$ at
  the LHC (in fb) as a function of $\sqrt{s}$ (in TeV) up to NNLO in
  QCD from Ref.~\cite{deFlorian:2013uza}, including the scale
  uncertainty. Center: the same at NLO using different PDF sets, from
  Ref.~\cite{Baglio:2012np}. Right: the same at NLO including the
  total theoretical uncertainty, from Ref.~\cite{Baglio:2012np}.}
\label{fig:ggHH}
\end{figure}

The gluon fusion mechanism provides the largest production channel. It
is mediated by loops of heavy quarks that are in the SM mainly top
quarks, see Fig.~\ref{fig:diagram} (up). The bottom loop contribution
amounts to less that 1\% at leading order (LO). The LO cross section
was calculated decades ago~\cite{Eboli:1987dy,Glover:1987nx} and the
process has been known for long at NLO in QCD in an EFT approach using
the infinite top quark mass approximation~\cite{Dawson:1998py}. The
NLO $K$-factor is of the order of 2, similar to the single Higgs
production case. The major improvement in 2013 came from the extension
of this calculation up to the NNLO order, providing a $+20\%$ increase
of the total cross section~\cite{deFlorian:2013uza}, and is depicted in
Fig.~\ref{fig:ggHH} (left). At 14 TeV one has
$\sigma^{\rm NLO}(gg\to HH) = 33.9$ fb and $\sigma^{\rm
  NNLO}(gg\to HH) = 40.2$
fb~\cite{deFlorian:2013uza}. A next-to-next-to-leading logarithmic
(NNLL) resummation was performed in Ref.~\cite{Shao:2013bz} and
increases the NLO cross section by $20\%$ to $30\%$, stabilizing also
the scale dependence of the result.

The gluon fusion channel is affected by sizable uncertainties, divided
in three categories: {\it a)} the scale uncertainty which is due to the
variation of the renormalization scale $\mu_R$ and the factorization
scale $\mu_F$ around a central scale $\mu_0=M_{HH}$, and viewed as a
rough estimate of the missing higher-order 
terms. This amounts to $\simeq \pm 18\%$ at NLO at 14
TeV~\cite{Baglio:2012np} and $\simeq \pm 8\%$ at
NNLO~\cite{deFlorian:2013uza}; {\it b)} the uncertainty related to the
parton distribution function (PDF) and the experimental value of 
$\alpha_s(M_Z^2)$, reflected in the spread of the predictions using
different PDF sets, see Fig.~\ref{fig:ggHH} (center). The uncertainty
calculated within the MSTW2008 PDF set~\cite{Martin:2009} at 90\% CL is
$\pm 7\%$ at 14 TeV~\cite{Baglio:2012np}; {\it c)} the uncertainty
related to the EFT approach (see Ref.~\cite{Baglio:2012np} for more
details), estimated to be of the order of 10\%~\cite{Baglio:2012np}
and confirmed by the top mass expansion calculation of
Ref.~\cite{Grigo:2013rya}. The total uncertainty amounts to $\pm 37\%$
at 14 TeV at NLO~\cite{Baglio:2012np}, which can be reduced down to
$\pm 30\%$ using the latest NNLO result.

\subsection{The vector boson fusion channel}

The VBF channel is the second largest production channel at the
LHC. The structure of this process is very similar to the single Higgs
production case and proceeds at LO via the generic Feynman diagrams
depicted in Fig.~\ref{fig:diagram} (down). The LO cross section has
been known for a while~\cite{Eboli:1987dy,Keung:1987nw} and recently
the NLO QCD corrections have been calculated for the total cross
section and the differential
distributions~\cite{Baglio:2012np,Frederix:2014hta} and they increase
the LO result by $\simeq 7\%$. The calculation has been implemented in
the public code {\tt VBFNLO}~\cite{Arnold:2008rz}. The approximate
NNLO QCD corrections have been obtained using the structure function 
approach which gives quite good results for the total cross section
and they increase the NLO result by less than
1\%~\cite{Liu-Sheng:2014gxa}.

The VBF channel is a rather clean process and the theoretical
uncertainties are rather small. The scale uncertainty, calculated with a
variation of $\mu_R$ and $\mu_F$ around the central scale
$\mu_0=Q^*_{W/Z}$ is roughly $ \pm 3\%$ at
NLO~\cite{Baglio:2012np}. The PDF uncertainty is limited and amounts
to $\simeq +7\% / - 4\%$ at 14 TeV. There is no EFT uncertainty and
the total theory error is $\simeq +8\% / -5\%$ at 14
TeV~\cite{Baglio:2012np}.

\subsection{Interface to parton shower}

The beginning of the year 2014 has seen progresses in the interface
between the hard cross section calculations and the parton showering
effects. Gluon fusion production plus one jet has been merged to
parton shower in an {\tt HERWIG++}
implementation~\cite{Bellm:2013lba}, leading to a sizable reduction of
the theoretical uncertainties on the efficiencies of the cut, now at
the level of $10\%$~\cite{Maierhofer:2013sha}.

In Ref.~\cite{Frederix:2014hta} a
NLO interface to parton shower was performed for all main processes in
the {\tt MadGraph5\_aMC@NLO} framework~\cite{Alwall:2014hca}, allowing
for NLO differentials predictions in all channels.
In particular, an improved NLO
calculation has been performed for the gluon fusion mechanism in which
the real emission is calculated exactly.

\section{Parton level analysis}

In order to extract the triple Higgs coupling $\lambda_{HHH}$, first
the Higgs pair production process needs to be measured. In 
Ref.~\cite{Baglio:2012np} it has been shown that the cross section
which is the most sensitive to $\lambda_{HHH}$ is the VBF
production. Increasing the center-of-mass energy reduces the
sensitivity of the total cross section to the triple Higgs coupling.
50\% accuracy on the total cross section leads to a 50\% accuracy on
$\lambda_{HHH}$ at 14 TeV.

Due to the smallness of the total cross sections, in the parton level
analyses it is required that at least one Higgs boson decays in a 
$b\bar{b}$ pair because this channel has the highest branching
fraction. There are then three interesting final states: {\it a)}
$b\bar{b} \tau\tau$; {\it b)} $b\bar{b} \gamma\gamma$, 
rather clean but the rates are very small and there is a lot of fake
photon identification; {\it c)} semi-leptonic $b\bar{b} W(\to \ell
\nu) W(\to 2j)$, rather difficult because of the missing energy. The
fully leptonic channel is not promising~\cite{Baglio:2012np} while the
other channels are currently used by the experimental collaborations
in their projections for the future~\cite{ATLAS:2013hta}. All the
analyses are based on the gluon fusion production channel at 14 TeV
using LO $gg\to HH$ matrix elements normalized to the NLO total cross
section and boosted topology cuts in addition to standard acceptance
cuts. The channel $HH+2j$, including VBF production, has started to be
investigated~\cite{Dolan:2013rja}.

\subsection{\boldmath The $b\bar{b}\tau\tau$ channel}

This channel is rather promising. When using a $\tau$ reconstruction
efficiency of 80\%, $M_{HH}> 350$ GeV and $p_T(H)>100$ GeV as boosted
topology cuts and an optimistic mass window $112.5$ GeV $ <
M_{\tau\tau} < 137.5$ GeV, this results in a significance
$S/\sqrt{B}=2.97$ already at 300 fb$^{-1}$
and 9.37 at 3 ab$^{-1}$~\cite{Baglio:2012np}. This corresponds
respectively to 33 and 330 signal events.

The major improvement has come from the use of the jet substructure
analysis~\cite{Butterworth:2008iy}. Defining a large cone size for the
jet (a ``fat jet'') and then working backward through the jet in order
to separate the softer subjets helps to distinguish the signal from
the large QCD backgrounds. This has been applied successfully in
addition to the cut strategy presented above and one obtains a
signal-over-background ratio $S/B\simeq 0.5$ and 95
signal events at 1000 fb$^{-1}$~\cite{Dolan:2012rv}. Adding one jet in
the final state enhances the significance and $S/B\simeq 1.5$, and
then with kinematic bounding variables a 60\% accuracy on the triple
Higgs coupling could be reached at 3
ab$^{-1}$~\cite{Barr:2013tda}. This is hence a very promising channel
that needs a dedicated analysis by the experimental collaborations as
these results represent what could be achieved in an ideal situation.

\subsection{\boldmath The $b\bar{b}\gamma\gamma$ channel}

The $b\bar{b}\gamma\gamma$ channel is a clean channel but rather
difficult because of the smallness of the signal rates and the large
amount of fake photons. Nevertheless it has been found in
Ref.~\cite{Baglio:2012np} that the significance could be
$S/\sqrt{B}=6.46$ at 3 ab$^{-1}$ with 47 signal events when assuming a
$b$-tagging efficiency of 70\% and simulating the fake photons with
{\tt DELPHES}~\cite{Ovyn:2009tx}. This simulation also uses the same
boosted topology cuts of the previous section with $|\eta_{H}|<2$ and
an isolation $\Delta R(b,b)<2.5$ in addition. This channel is then
also very promising and it has been part of a high energy LHC
analysis~\cite{Yao:2013ika}.

Using a multivariate analysis could improve the results. It was found in
Ref.~\cite{Barger:2013jfa} that it increases the significance of the
signal and would lead to a probe of the triple Higgs coupling at the
level of $40\%$ uncertainty at the LHC at 14 TeV using 3 ab$^{-1}$ of
data.

\subsection{The semi-leptonic \boldmath $b\bar{b}W^+W^-$ channel}

Whereas the fully leptonic $b\bar{b}W^+W^-$ channel seems to be
hopeless, the semi-leptonic channel could trigger interesting
results. In Ref.~\cite{Papaefstathiou:2012qe} a parton level analysis
was presented, that relies on a jet substructure analysis improved
with a boosted decision tree and specific cuts to this channel such as
a cut on the hadronically decaying $W$ boson $m_{W_h}>65$ GeV. The
analysis has obtained a promising result of $S/\sqrt{S+B} = 2.4$ at
600 fb$^{-1}$ already with 9 signal events. More detailed analyses are
required to assess the potential of this search channel in a more
realistic experimental environment.

\subsection{More improvements}

There are additional improvements that can increase the sensitivity of
the previous searches. One example is the use of ratios $C_{HH}$ of
double Higgs production to single Higgs production cross
sections~\cite{Goertz:2013kp}. Owing to the similar structure in the
higher-order corrections in both channels, this leads to a substantial
reduction of the theoretical uncertainties with $\Delta^{\mu} C_{HH}
\simeq \pm 2\%$, $\Delta^{\rm PDF}C_{HH} \simeq \pm 2\%$. This would
lead to a very promising confidence interval of $\simeq +30\% / -20\%$
on the triple Higgs coupling when combining the three previous search
channels.

An analysis in the 4$b$ search channel, which had
been thought for long not a useful channel, has been recently
released. Using a jet substructure analysis and a side-band analysis, it
was found that at the LHC at 14 TeV with 3 ab$^{-1}$ of data it may be
possible to constraint $\lambda_{HHH} < 1.2 \times \lambda_{HHH}^{\rm
  SM}$  at 95\% CL~\cite{deLima:2014dta}. More experimental analyses
are obviously required to confirm this result.

\section{Outlook}

The production of a Higgs boson pair is one of the goals of
the high luminosity run of the LHC at 14 TeV, in order to extract the
triple Higgs coupling. The past two years have seen major improvements
in the theoretical knowledge on the SM Higgs boson pair production and
the main channels have now reached the NLO or even NNLO QCD
accuracy. The theoretical uncertainty is of the order of 30\% in the
gluon fusion channel and less than 10\% in the other production
channels.

The parton level analyses, notably in the $b\bar{b}\tau\tau$ and
$b\bar{b}\gamma\gamma$ channels, have seen good prospects 
already at 300 fb$^{-1}$ and mostly at 3 ab$^{-1}$, triggering the
experimental collaborations to perform a detailed study. Major
theoretical improvements are expected in the coming years towards a
fully differential NLO calculation of the gluon fusion channel
including the full quark mass dependance.

\acknowledgments{
JB would like to thank the organizers for the invitation and
acknowledges the support from the DFG under the SFB TR-9
Computational Particle Physics.}


\begin{thebibliography}{99}
\bibitem{Aad:2012tfa}
  G.~Aad {\it et al.}  [ATLAS Collaboration],
  Phys.\ Lett.\ B {\bf 716} (2012) 1;
  S.~Chatrchyan {\it et al.}  [CMS Collaboration],
  Phys.\ Lett.\ B {\bf 716} (2012) 30.

\bibitem{Englert:1964et}
  F.~Englert and R.~Brout,
  Phys.\ Rev.\ Lett.\  {\bf 13} (1964) 321;
  P.~W.~Higgs,
  Phys.\ Lett.\  {\bf 12} (1964) 132; 
  Phys.\ Rev.\ Lett.\  {\bf 13} (1964) 508;
  G.~S.~Guralnik, C.~R.~Hagen and T.~W.~B.~Kibble,
  Phys.\ Rev.\ Lett.\  {\bf 13} (1964) 585.  

\bibitem{ATLAS:conf}
  ATLAS Collaboration,
  ATLAS-CONF-2013-034.
  S.~Chatrchyan {\it et al.}  [CMS Collaboration],
  JHEP {\bf 1306} (2013) 081.
  
\bibitem{Plehn:2005nk}
  T.~Plehn and M.~Rauch,
  Phys.\ Rev.\ D {\bf 72} (2005) 053008.

\bibitem{Boudjema:1995cb}
  F.~Boudjema and E.~Chopin,
  Z.\ Phys.\ C {\bf 73} (1996) 85.

\bibitem{Djouadi:1999gv}
  A.~Djouadi, W.~Kilian, M.~M\"{u}hlleitner and P.~M.~Zerwas,
  Eur.\ Phys.\ J.\ C {\bf 10} (1999) 27.

\bibitem{Djouadi:1999rca}
  A.~Djouadi, W.~Kilian, M.~M\"{u}hlleitner and P.~M.~Zerwas,
  Eur.\ Phys.\ J.\ C {\bf 10} (1999) 45.

\bibitem{Baur:2002rb}
  U.~Baur, T.~Plehn and D.~L.~Rainwater,
  Phys.\ Rev.\ Lett.\  {\bf 89} (2002) 151801;
  Phys.\ Rev.\ D {\bf 67} (2003) 033003;
  Phys.\ Rev.\ D {\bf 69} (2004) 053004.




\bibitem{deFlorian:2013uza}
  D.~de Florian and J.~Mazzitelli,
  Phys.\ Lett.\ B {\bf 724} (2013) 306;
  Phys.\ Rev.\ Lett.\  {\bf 111} (2013) 201801.

\bibitem{Baglio:2012np}
  J.~Baglio, A.~Djouadi, R.~Gr\"{o}ber, M.~M.~M\"{u}hlleitner, J.~Quevillon and M.~Spira,
  JHEP {\bf 1304} (2013) 151.

\bibitem{Frederix:2014hta}
  R.~Frederix, S.~Frixione, V.~Hirschi, F.~Maltoni, O.~Mattelaer, P.~Torrielli, E.~Vryonidou and M.~Zaro,
  Phys.\ Lett.\ B {\bf 732} (2014) 142.

\bibitem{Liu-Sheng:2014gxa}
  L.~Liu-Sheng, Z.~Ren-You, M.~Wen-Gan, G.~Lei, L.~Wei-Hua and L.~Xiao-Zhou,
  Phys.\ Rev.\ D {\bf 89} (2014) 073001.

\bibitem{Eboli:1987dy}
  O.~J.~P.~Eboli, G.~C.~Marques, S.~F.~Novaes and A.~A.~Natale,
  Phys.\ Lett.\ B {\bf 197} (1987) 269.

\bibitem{Glover:1987nx}
  E.~W.~N.~Glover and J.~J.~van der Bij,
  Nucl.\ Phys.\ B {\bf 309} (1988) 282;
  D.~A.~Dicus, C.~Kao and S.~S.~D.~Willenbrock,
  Phys.\ Lett.\ B {\bf 203} (1988) 457;
  T.~Plehn, M.~Spira and P.~M.~Zerwas,
  Nucl.\ Phys.\ B {\bf 479} (1996) 46
  [Erratum-ibid.\ B {\bf 531} (1998) 655].
  
\bibitem{Dawson:1998py}
  S.~Dawson, S.~Dittmaier and M.~Spira,
  Phys.\ Rev.\ D {\bf 58} (1998) 115012.

\bibitem{Shao:2013bz}
  D.~Y.~Shao, C.~S.~Li, H.~T.~Li and J.~Wang,
  JHEP {\bf 1307} (2013) 169.

\bibitem{Martin:2009}
  A.~Martin, W.~Stirling, R.~Thorne and G.~Watt,
  Eur.\ Phys.\ J.\ C {\bf 63} (2009) 189.

\bibitem{Grigo:2013rya}
  J.~Grigo, J.~Hoff, K.~Melnikov and M.~Steinhauser,
  Nucl.\ Phys.\ B {\bf 875} (2013) 1.

\bibitem{Keung:1987nw}
  W.~-Y.~Keung,
  Mod.\ Phys.\ Lett.\ A {\bf 2} (1987) 765;
  D.~A.~Dicus, K.~J.~Kallianpur and S.~S.~D.~Willenbrock,
  Phys.\ Lett.\ B {\bf 200} (1988) 187;
  A.~Dobrovolskaya and V.~Novikov,
  Z.\ Phys.\ C {\bf 52} (1991) 427.

\bibitem{Arnold:2008rz}
  K.~Arnold, M.~B\"{a}hr, G.~Bozzi, F.~Campanario, C.~Englert, T.~Figy, N.~Greiner and C.~Hackstein {\it et al.},
  Comput.\ Phys.\ Commun.\  {\bf 180} (2009) 1661;
  J.~Baglio, J.~Bellm, F.~Campanario, B.~Feigl, J.~Frank, T.~Figy, M.~Kerner and L.~D.~Ninh {\it et al.},
  arXiv:1404.3940 [hep-ph].

\bibitem{Bellm:2013lba}
  J.~Bellm, S.~Gieseke, D.~Grellscheid, A.~Papaefstathiou, S.~Pl\"{a}tzer, P.~Richardson, C.~R\"{o}hr and T.~Schuh {\it et al.},
  arXiv:1310.6877 [hep-ph].

\bibitem{Maierhofer:2013sha}
  P.~Maierh\"{o}fer and A.~Papaefstathiou,
  JHEP {\bf 1403} (2014) 126.

\bibitem{Alwall:2014hca}
  J.~Alwall, R.~Frederix, S.~Frixione, V.~Hirschi, F.~Maltoni, O.~Mattelaer, H.~-S.~Shao and T.~Stelzer {\it et al.},
  arXiv:1405.0301 [hep-ph].

\bibitem{ATLAS:2013hta}
  ATLAS Collaboration,
  arXiv:1307.7292 [hep-ex];
  CMS Collaboration,
  arXiv:1307.7135 [hep-ex].

\bibitem{Dolan:2013rja}
  M.~J.~Dolan, C.~Englert, N.~Greiner and M.~Spannowsky,
  Phys.\ Rev.\ Lett.\  {\bf 112} (2014) 101802.

\bibitem{Butterworth:2008iy}
  J.~M.~Butterworth, A.~R.~Davison, M.~Rubin and G.~P.~Salam,
  Phys.\ Rev.\ Lett.\  {\bf 100} (2008) 242001.

\bibitem{Dolan:2012rv}
  M.~J.~Dolan, C.~Englert and M.~Spannowsky,
  JHEP {\bf 1210} (2012) 112.

\bibitem{Barr:2013tda}
  A.~J.~Barr, M.~J.~Dolan, C.~Englert and M.~Spannowsky,
  Phys.\ Lett.\ B {\bf 728} (2014) 308.

\bibitem{Ovyn:2009tx}
  S.~Ovyn, X.~Rouby and V.~Lemaitre,
  arXiv:0903.2225 [hep-ph].

\bibitem{Yao:2013ika}
  W.~Yao,
  arXiv:1308.6302 [hep-ph].

\bibitem{Barger:2013jfa}
  V.~Barger, L.~L.~Everett, C.~B.~Jackson and G.~Shaughnessy,
  Phys.\ Lett.\ B {\bf 728} (2014) 433.

\bibitem{Papaefstathiou:2012qe}
  A.~Papaefstathiou, L.~L.~Yang and J.~Zurita,
  Phys.\ Rev.\ D {\bf 87} (2013) 011301.

\bibitem{Goertz:2013kp}
  F.~Goertz, A.~Papaefstathiou, L.~L.~Yang and J.~Zurita,
  JHEP {\bf 1306} (2013) 016.

\bibitem{deLima:2014dta}
  D.~E.~Ferreira de Lima, A.~Papaefstathiou and M.~Spannowsky,
  arXiv:1404.7139 [hep-ph].

\end{thebibliography}
\end{document}